\documentclass[11pt,a4paper]{article}
\usepackage{jcappub}
\usepackage{amsmath}
\usepackage{amssymb}
\usepackage{latexsym}
\usepackage{graphicx}
\usepackage{subfig}
\usepackage{hyperref}
\usepackage{mathrsfs}
\usepackage{color}
\usepackage{bm}
\usepackage{url}
\usepackage{hyperref}
\usepackage{epstopdf}
\newcommand{\be}{\begin{equation}}
\newcommand{\ee}{\end{equation}}
\def\bea{\begin{eqnarray}}
\def\eea{\end{eqnarray}}
\def\ba{\begin{eqnarray}}
\def\ea{\end{eqnarray}}

\begin{document}

\title{Reconstruction of dark energy and expansion dynamics using Gaussian processes}
\author[a]{Marina Seikel}
\author[a]{Chris Clarkson}
\author[a,b]{and Mathew Smith}

\affiliation[a]{Astrophysics, Cosmology and Gravity Centre (ACGC), and Department of Mathematics and Applied Mathematics, University of Cape Town, Rondebosch, 7701, SA}
\affiliation[b]{Department of Physics, University of the Western Cape, Bellville, 7535, Cape Town, SA}

\emailAdd{marina.seikel@uct.ac.za}
\emailAdd{chris.clarkson@uct.ac.za}
\emailAdd{mathew.smith@uct.ac.za}

\arxivnumber{1204.2832}

\abstract{
An important issue in cosmology is reconstructing the effective dark
energy equation of state directly from observations.  With few
physically motivated models, future dark energy studies cannot only be
based on constraining a dark energy parameter space, as the errors
found depend strongly on the parametrisation considered. We present a
new non-parametric approach to reconstructing the history of the
expansion rate and dark energy using Gaussian Processes, which is a
fully Bayesian approach for smoothing data. We present a pedagogical
introduction to Gaussian Processes, and discuss how it can be used to
robustly differentiate data in a suitable way.  Using this method we
show that the Dark Energy Survey - Supernova Survey (DES) can
accurately recover a slowly evolving equation of state to $\sigma_w =
\pm0.05$ ($95\%$ CL) at $z=0$ and $\pm0.25$ at $z=0.7$, with a minimum
error of $\pm0.025$ at the sweet-spot at $z\sim0.16$, provided the
other parameters of the model
are known. Errors on the expansion history are an order of magnitude
smaller, yet make no assumptions about dark energy whatsoever. A code
for calculating functions and their first three derivatives using
Gaussian processes has been developed and is available for
\href{http://www.acgc.uct.ac.za/~seikel/GAPP/index.html}{\sl
  download}.} 

\maketitle

\section{Introduction}


A key problem in cosmology lies in determining whether dark energy is a cosmological constant and if not, then constraining how it evolves with cosmic time. Previously, this has been approached in a model-building way~-- i.e., constraining specific models of dark energy, such as quintessence models or modifications to general relativity~\cite{copeland}. It has been a significant problem to produce well motivated models which are not ad hoc in some way. In this sense they tend to have functional degrees of freedom~-- such as the quintessence potential or the `$f$' in $f(R)$ theories of gravity~-- which have to be constrained via observations. Constraints on dark energy are currently derived after free functions in the models are parametrised in simple ways. 

Alternatively, one can approach the problem in a different way and look for any deviations from a cosmological constant, irrespective of origin. 
The dark energy equation of state $w(z)=p(z)/\rho(z)$ is typically constrained  using
distance measurements as a function of redshift. 
The luminosity distance may be written as 
\be
d_{L}(z)=\frac{c(1+z)}{H_0 \sqrt{-\Omega_k}}\sin{\left( 
\sqrt{-\Omega_k}\int_0^z{\mathrm{d}z'\frac{H_0}{H(z')}}\right)},
\label{lum_d}
\ee
where $H(z)$ is given by the Friedmann equation, 
\be
H(z)^2 =  H_0^2\left\{\Omega_{m} (1+z)^3+\Omega_{k}(1+z)^2
+(1-\Omega_m-\Omega_k)\exp{\left[3\int_0^z
\frac{1+w(z')}{1+z'}\mathrm{d}z'\right]}\right\},
\label{hubble}
\ee
where $H_0=H(z=0)$ and $\Omega_{m,k}$ are the normalised density
parameters. Without a model for dark energy it is difficult to use other observations as these require perturbations at some level, so distances are the vital observable. 
A common procedure here is to postulate a multiple parameter
form for $w(z)$ and calculate $d_L(z)$. The most promising of these
approaches uses a principal component analysis to construct the
`optimal' basis functions for $w(z)$ based on the data
available~\cite{HS,Crittenden:2005wj,Crittenden:2011aa}. Another uses Gaussian Processes to effectively smooth $w$ and fit it to data~\cite{Holsclaw10} (we discuss this work below).
 
An alternative method is to reconstruct $w(z)$ by directly
reconstructing the luminosity-distance curve.  Writing
$D(z)=(H_0/c)(1+z)^{-1}d_L(z)$ as the normalised comoving distance, we
have~\cite{Starobinsky:1998fr,Nakamura:1998mt,Huterer:1998qv,saini}:
\be\label{w}
w(z)= \frac{2(1+z)(1+\Omega_kD^2)D''-[(1+z)^2\Omega_kD'^2
+2(1+z)\Omega_kDD'-3(1+\Omega_kD^2)]D'}{3\{(1+z)^2[\Omega_k+(1+z)\Omega_m]D'^2-(1+\Omega_kD^2)\}D'}
\ee
Thus, given a distance-redshift curve $D(z)$, we can reconstruct the
dark energy equation of state, assuming we know the density parameters
$\Omega_m$ and $\Omega_k$. Different methods for doing this involve
smoothing the data to give $D(z)$, or parametrising $D(z)$ by a
function; see~\cite{Sahni:2006pa} for a comprehensive review,
and~\cite{weller_albrecht,Alam:2003sc,daly,Alam:2004jy,Wang:2004py,Daly:2004gf,shafieloo,Clarkson:2010bm,Lazkoz:2012eh}
for alternative model independent approaches.

The direct reconstruction method is unstable because of the two
derivatives of the observed function in Eq.~\eqref{w}, requiring the
fitting function to accurately capture the slope and concavity of
luminosity distance curve.  This means that differences between the
true underlying model and the fitted function due to the choice of
parametrisation are amplified drastically when reconstructing~$w$. Furthermore, $w$ is constructed from a quotient of functions
which need to balance to obtain the correct $w$; the denominator can
easily pass through zero for even small uncertainties making it
especially unstable. Typical reconstruction methods usually appear to flounder
even at moderate redshift for this reason. However, such problems may be considered as informing us about the real errors on $w$ without the intrinsic priors which arise when we start from a parametric form of $w$ and constrain it from there. 

\begin{figure}
\subfloat[$\Omega_m=0.275\pm 0.016$]{
\includegraphics[width=0.5\textwidth]{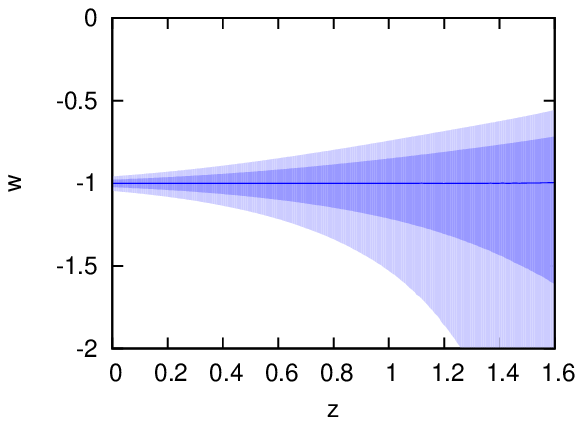}}
\subfloat[$\Omega_k=0.00 \pm 0.01$]{
\includegraphics[width=0.5\textwidth]{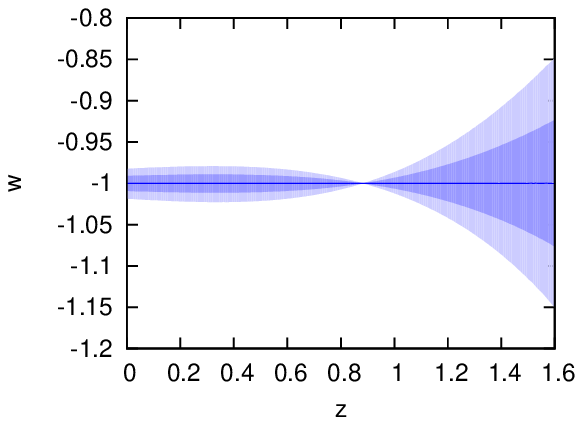}}
\caption{ {\em Left:} Dark energy reconstruction using Eq.~\eqref{w} for
  $\Omega_k=0$, when $D(z)$ and its derivatives are exactly known
  ($\Lambda$CDM with $\Omega_m=0.275$). The
  uncertainty in the reconstructed $w$ (the blue shaded regions show
  the 68\% and 95\% CL) only comes from the prior on the
  matter density, $\Omega_m=0.275\pm 0.016$ (WMAP7 \cite{Komatsu}). 
  {\em Right:} The same plot for fixed matter density,
  $\Omega_m=0.275$, and a prior on the curvature of $\Omega_k=0.00 \pm 0.01$.}
\label{exactD}
\end{figure}

Indeed, even small errors in the parameters can lead to large errors
in $w(z)$. This can be seen in the left panel of Fig.~\ref{exactD}, where we have
assumed that we know $D(z)$, $D'(z)$, $D''(z)$ and $\Omega_k$ exactly,
and have used the WMAP7 constraints on the matter density,
$\Omega_m=0.275\pm 0.016$~\cite{Komatsu}. A similar effect happens from uncertainties in the curvature. Even under these idealised
conditions, the errors of the reconstructed $w(z)$ are large because
they properly take into account the degeneracies with the density
parameters~\cite{Clarkson:2007bc,Kunz:2007rk,Hlozek:2008mt,Barenboim:2009ug}.

Nevertheless, such approaches play an important role in our
understanding of dark energy for several reasons. A simple one is that
$w(z)$ may only be a place-holder phenomenological function supposed
to encapsulate all possible alternatives to dark energy, such as
modified gravity~\cite{Durrer:2008in}, and not just simple
quintessence models. This means that we have to accept there could be
really unexpected behaviour.  An important example of this happens if $\rho_{\mathrm{eff}}$
passes through zero at some $z$ while $p_\mathrm{eff}$ remains
non-zero, then $w(z)$ has a pole at $z$. To evaluate the integral
appearing in $H(z)$ requires integration around the pole, and the
residue of the integral taken into account (assuming it is defined). No method which starts
from a set of basis functions for $w(z)$ can recover this behaviour if
it is not known \emph{a priori}.

Consider the explanation for dark energy suggested by causal set
theory~\cite{Ahmed:2002mj} wherein we should expect the value of
$\Lambda$ to stochastically fluctuate around the Hubble scale. In that
case, $w$ would be discontinuous and varying widely over short redshift scales, but this can never be 
represented by choosing simple functions in $w$-space.

Because of these reasons, methods which start from $w$ and work towards $D(z)$ can
underestimate the errors in our understanding of dark energy. The
errors which appear to condemn reconstruction methods which start from
$D(z)$ and work towards $w$ are actually much more representative of
our genuine errors in this regard. It is for this reason that it is
important to pursue reconstruction methods, even though they are very
challenging.

In this work, we use Gaussian processes for the reconstruction of
$D(z)$ and its derivatives. 
Then equation \eqref{w} is used to determine $w(z)$. 
Gaussian processes describe a distribution over functions and are thus
a generalisation of Gaussian distributions to function space. The
analysis is fully Bayesian; we start with a
prior for the function distribution and combine it with the likelihood
of observing the data, given that distribution. This leads to a
posterior function distribution.

Gaussian processes in combination with MCMC methods have previously been
used to reconstruct $w(z)$ by Holsclaw {\em et
  al.}~\cite{Holsclaw10,Holsclaw11}. While their method uses
integration over $w(z)$ to obtain the distance, we reconstruct $D(z)$
and its derivatives in order to determine $w(z)$. Gaussian processes
typically have an implicit prior favoring smooth functions. This is closely
related to the preference of simpler models in Bayesian model selection. So in our approach we have a smoothness prior on our distance data, whereas, by applying the GP to $w(z)$ directly the smoothness prior is rather different in~\cite{Holsclaw10,Holsclaw11}. As we shall see we find  different results at high redshift. 
A comparison of the
two methods will be given in Section~\ref{discussion}.

The outline of the paper is as follows:
We start with an introduction to Gaussian processes, including a
performance test for the reconstruction of a function and its
derivatives. In Section~\ref{reconstruction}, Gaussian processes are
used to reconstruct $H(z)$, $q(z)$ and $w(z)$ for a mock SN data set
and for the Union2.1 data set \cite{Suzuki}. The results are discussed in
Section~\ref{discussion}. 

\section{Gaussian Processes}

In this section, we summarise the Gaussian process algorithm, which
can perform a reconstruction of a function from data without assuming
a parametrisation of the function. We mainly follow the book by
Rasmussen and Williams~\cite{Rasmussen06}. Other useful references may
be found in \cite{MacKay,Williams} and on the Gaussian Process webpage
\cite{GPorg}.
We have developed a code for Gaussian processes called \href{http://www.acgc.uct.ac.za/~seikel/GAPP/index.html}{GaPP} (Gaussian
Processes in Python) which is available for
download\footnote{\href{http://www.acgc.uct.ac.za/~seikel/GAPP/index.html}{\url{http://www.acgc.uct.ac.za/~seikel/GAPP/index.html}}}.
It can be used to reconstruct a function and its derivatives, from a given data set.

Given a data set $\mathcal{D}$ of $n$ observations: 
\be
\mathcal{D} =
\{ (x_i,y_i)| i=1,\dots , n \}
\ee
we would like to reconstruct a
function $f(x)$ that describes the data. We write the set of training
inputs, i.e.\ the locations $\{x_i\}_{i=1}^n$ of the observations, as
$\bm X$. The locations, at which we want to reconstruct the function, are
denoted as $\bm X^*$.

\subsection{Reconstructing a function}

A Gaussian process is the generalisation of a Gaussian
distribution. While the latter is the distribution of a random
variable, the Gaussian process describes a distribution over
functions. Consider a function $f$ formed from a Gaussian process. The
value of $f$ when evaluated at a point $x$ is a Gaussian random
variable with mean $\mu(x)$ and variance $\text{Var}(x)$. The function
value at $x$ is not independent of the function value at some other
point $\tilde{x}$ (especially when $x$ and $\tilde{x}$ are close to
each other), but is related by a covariance function
$\text{cov}\left( f(x),f(\tilde{x}) \right) = k(x,\tilde{x})$.  Thus,
the distribution of functions can be described by the following
quantities:
\begin{eqnarray}
\mu(x) &=& \mathbb{E}[f(x)]\,,\\
k(x,\tilde{x}) &=& \mathbb{E}[(f(x)-\mu(x))(f(\tilde{x})-\mu(\tilde{x}))]\,,\\  
\text{Var}(x) &=& k(x,x)\,.
\end{eqnarray}
The Gaussian process is written as
\begin{equation}
f(x) \sim \mathcal{GP}\left( \mu(x), k(x,\tilde{x})
\right) \;.
\end{equation}

There is a wide range of possible covariance functions. While one will
often chose covariance functions that only depend on the distance
between the input points $|x-\tilde{x}|$, this is not a necessary requirement.
Throughout this work, we use the squared exponential covariance
function:
\begin{equation}
k(x,\tilde{x}) =
\sigma_f^2 \exp\left( -\frac{(x - \tilde{x})^2}{2\ell^2} \right) \;.
\end{equation}
This function has the advantage that it is infinitely
differentiable, which is useful for reconstructing the
derivative of a function.
The squared exponential covariance function depends on the two
`hyperparameters' $\sigma_f$ and $\ell$. In contrast to actual
parameters, the hyperparameters do not specify the form of a
function. Instead they characterize the ``bumpiness'' of
the function. The characteristic length scale
$\ell$ can be thought of as the distance one has to travel in
$x$-direction to get a significant change in $f(x)$, whereas the
signal variance $\sigma_f$ denotes the typical change in the
$y$-direction.

For a set of input points $\bm X=\{x_i\}$, the covariance matrix
$K(\bm X,\bm X)$
is given by $[K(\bm X,\bm X)]_{ij}=k(x_i,x_j)$.
Even without any observations, one can use the covariance matrix to
generate a random function $f(x)$ from the Gaussian process, i.e.\ one
generates a Gaussian vector ${ \bm f^*}$ of function values at
$\bm X^*$ with $f^*_i=f(x^*_i)$:
\begin{equation}\label{prior}
{\bm f^*} \sim \mathcal{N}\left( \text{\boldmath $\mu^*$},K(\bm X^*,\bm X^*)
\right) \;,
\end{equation}
where {\boldmath $\mu^*$} is the {\em a priori} assumed mean of ${\bm
  f^*}$. The notation $\mathcal{N}$ means the Gaussian process
$\mathcal{GP}$ is evaluated at specific points $x^*$, where $f(x^*)$
is a random value drawn from a normal distribution.  As the function
is not restricted by any observations, it is quite arbitrary. Its
values at different locations $x^*_i$ are however correlated by the
covariance function. This can be considered as a prior on the choice
of output functions; only when we add in data at other points $x_i$
does the output become constrained further.

Observational data $(x_i,y_i)$ can also be described by a Gaussian
process, assuming the errors are Gaussian. The actual observations are
assumed to be scattered around the underlying function, i.e.\ $y_i =
f(x_i) + \epsilon_i$, where Gaussian noise $\epsilon_i$ with variance
$\sigma_i^2$ is assumed. This variance has to be added to the
covariance matrix:
\be
{\bm y} \sim \mathcal{N}\left( \text{\boldmath $\mu$},K(\bm X,\bm X) +
C
\right)\,, 
\ee
where $C$ is the covariance matrix of the data. For uncorrelated data
we have simply $C=\text{diag}(\sigma_i^2)$.
The two Gaussian processes for ${\bm
  f^*}$ and ${\bm y}$ can be combined in the joint distribution:
\begin{equation}
\begin{bmatrix} {\bm y} \\ {\bm f^*} \end{bmatrix} \sim
\mathcal{N} \left( 
\begin{bmatrix}
\text{\boldmath $\mu$}\\
\text{\boldmath $\mu^*$}
\end{bmatrix}
,
\begin{bmatrix}
K(\bm X,\bm X) + C & K(\bm X,\bm X^*)\\
K(\bm X^*,\bm X)   & K(\bm X^*,\bm X^*)
\end{bmatrix}
\right)
\end{equation}

While ${\bm y}$ is known from observations, we want to reconstruct
${\bm f^*}$. This can be done with the conditional distribution (see the Appendix)
\begin{equation}\label{post}
{\bm f^*}| \bm X^*, \bm X,\bm y \sim \mathcal{N} \left(
\overline{\bm f^*}, \text{cov}({\bm f^*})
\right) \;,
\end{equation}
where
\begin{equation}\label{gp-mean}
\overline{\bm f^*} = \text{\boldmath $\mu^*$} + K(\bm X^*,\bm X)\left[K(\bm X,\bm X) +
  C\right]^{-1} ({\bm y} -  \text{\boldmath $\mu$})
\end{equation}
and
\begin{equation}
\text{cov}({\bm f^*}) = K(\bm X^*,\bm X^*) 
- K(\bm X^*,\bm X) \left[K(\bm X,\bm X) +
  C\right]^{-1}  K(\bm X,\bm X^*) \label{gp-cov}
\end{equation}
are the mean and covariance of ${\bm f^*}$, respectively. The variance
of ${\bm f^*}$ is simply the diagonal of $\text{cov}({\bm
  f^*})$. Eq.~(\ref{post}) is the posterior distribution of the
function given the data and the prior Eq.~(\ref{prior}). 

In order to be able to use the above equations for reconstructing a
function, we still need to know the hyperparameters $\sigma_f$ and
$\ell$. They can be trained by maximizing the marginal likelihood,
which is the marginalization over function values ${\bm f}$ at
locations $\bm X$:
\begin{equation}\label{marginal-p}
p({\bm y}|\bm X,\sigma_f,\ell) = \int p({\bm y}|{\bm
  f},\bm X)p({\bm f}|\bm X,\sigma_f,\ell) d{\bm f}\;.
\end{equation}
Note that the marginal likelihood only depends on the locations $\bm X$ of
the observations, but {\em not} on the points $\bm X^*$, where we want to
reconstruct the function.

With a Gaussian prior ${\bm f}|\bm X,\sigma_f,\ell \sim
\mathcal{N}(\text{\boldmath $\mu$},K(\bm X,\bm X))$ and with ${\bm y}|{\bm f}\sim
\mathcal{N}({\bm f},C)$, the integration of
\eqref{marginal-p} yields the log marginal likelihood
\begin{eqnarray}\label{log-marginal-p}
\ln \mathcal{L} &=& \ln p({\bm y}|\bm X,\sigma_f,\ell) \nonumber\\
&=& -\frac{1}{2}({\bm
  y} - \text{\boldmath $\mu$})^T \left[K(\bm X,\bm X) +
  C\right]^{-1} ({\bm y} - \text{\boldmath
  $\mu$})  - \frac{1}{2}\ln
  \left|K(\bm X,\bm X)+C\right|
  - \frac{n}{2}\ln 2\pi  \;.
\end{eqnarray}
The hyperparameters $\sigma_f$ and $\ell$ can now be optimized by
maximizing equation \eqref{log-marginal-p}.

In a completely Bayesian analysis, one should marginalise over the
hyperparameters instead of optimising them. This can be done with MCMC
methods. However in most cases, the log marginal likelihood is sharply
peaked. Thus, the optimisation is a very good approximation of the
marginalisation.

\begin{figure}
\subfloat[prior]{
\includegraphics[width=0.5\textwidth]{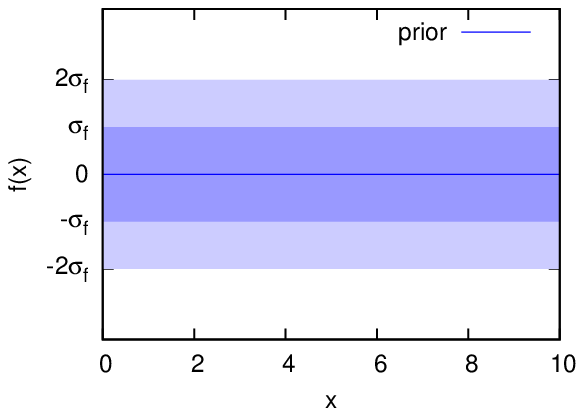}}
\subfloat[posterior]{
\includegraphics[width=0.5\textwidth]{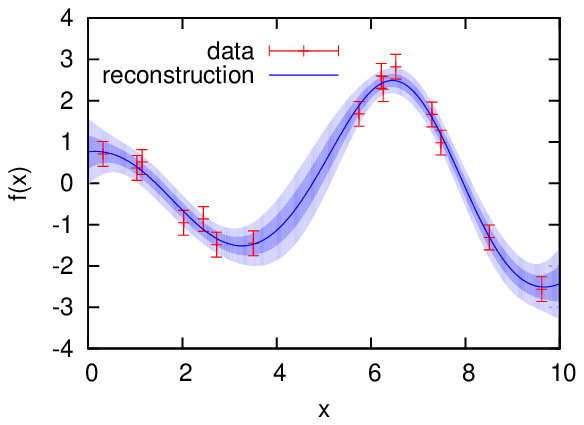}}
\caption{ {\em Left:} Prior of the Gaussian process as given by
  Eq.~\eqref{prior}. Note that the hyperparameters have not been
  trained yet.
  {\em Right:} Posterior of the Gaussian process as given by
  Eq.~\eqref{post}.}
\label{prior_post}
\end{figure}

Figure~\ref{prior_post} shows an example for the prior and the
posterior of a Gaussian process, where we have assumed a zero {\em a
  priori} mean function $\mu(x)=0$. For the prior, the hyperparameters
have not been trained yet. Thus, the scale of the $y$-axis is not
fixed and all functions are still possible. Adding data constrains the
function space, which can be seen in the plot for the posterior.

\subsection*{How to GP}

The key steps involved in constructing the $\mathcal{GP}$ are:
\begin{itemize}
\item Choose your data set.
\item Choose a covariance function. 
\item Choose points $x^*_i$ where we want the function $f^*_i$ to be estimated.
\item Decide a prior mean function $\mu(x)$. $\mu=\,$const. is a safe choice.
\item Train the hyperparameters:
\begin{itemize}
\item Carefully choose initial values for the hyperparameters. It is
  recommended to try different initial values because sometimes the
  optimisation of the hyperparameters can get stuck in a local maximum.
\item Maximise the likelihood function, Eq.~(\ref{log-marginal-p}), for the hyperparameters.
\end{itemize}
\item Calculate $\overline{\bm f^*}$ from Eq.~(\ref{gp-mean}). This gives the expected value function.
\item Calculate the
  diagonal elements of $\text{cov}(\bm f^*)$, Eq.~(\ref{gp-cov}).
  This gives the variance of $\bm f^*$.  
\end{itemize}

\subsection{Reconstructing the derivative of a function}

This method can also be used to reconstruct derivatives of $f(x)$ as 
the derivative of a Gaussian process is again a Gaussian
process. While the covariance between the observational points stays
the same, one also needs a covariance between the function and its
derivative and one between the derivatives for the
reconstruction. These covariances can be obtained by differentiating
the original covariance function:
\begin{equation}
\text{cov}\left(f_i,\frac{\partial f_j}{\partial x_j} \right) = 
\frac{\partial k(x_i,x_j)}{\partial x_j}
\ee
and
\be
\text{cov}\left(\frac{\partial f_i}{\partial x_i}, \frac{\partial
  f_j}{\partial x_j} \right) = 
\frac{\partial^2 k(x_i,x_j)}{\partial x_i\partial x_j}
\end{equation}
Covariances for higher derivatives of $f$ are calculated analogously.

Given the Gaussian process for $f(x)$, the Gaussian processes for the
first and second derivative are consequently given by:
\begin{eqnarray}
f(x) &\sim& \mathcal{GP}\left( \mu(x), k(x,\tilde{x})\right)\\
f'(x) &\sim& \mathcal{GP}\left( \mu'(x),
\frac{\partial^2k(x,\tilde{x})}{\partial x\,\partial\tilde{x}} \right)\\
f''(x) &\sim& \mathcal{GP}\left( \mu''(x),
\frac{\partial^4k(x,\tilde{x})}{\partial x^2\,\partial\tilde{x}^2} \right)
\end{eqnarray}

In the following, we only show the procedure for reconstructing the first
derivative of $f$. Reconstructions of higher derivatives are done
analogously. The joint distribution of ${\bm y}$ and ${\bm f^*}'$ is
\begin{equation}
\begin{bmatrix} {\bm y} \\ {\bm f^*}' \end{bmatrix} \sim
\mathcal{N} \left( 
\begin{bmatrix}
\text{\boldmath $\mu$}\\
\text{\boldmath $\mu^*$}'
\end{bmatrix}
,  
\begin{bmatrix}
K(\bm X,\bm X) + C & K'(\bm X,\bm X^*)\\
K'(\bm X^*,\bm X)              & K''(\bm X^*,\bm X^*)
\end{bmatrix}
\right)\;,
\end{equation}
where
\begin{equation}
[K'(\bm X,\bm X^*)]_{ij} = \frac{\partial k(x_i,x^*_j)}{\partial x^*_j}
\ee
and
\be
[K''(\bm X^*,\bm X^*)]_{ij} = \frac{\partial^2k(x^*_i,x^*_j)}{\partial
    x^*_i\,\partial x^*_j} \;.
\end{equation}
$K'(\bm X^*,\bm X)$ is the transpose of $K'(\bm X,\bm X^*)$.

Then the conditional distribution is given by:
\begin{equation}
{\bm f^*}'| \bm X^*, \bm X,y \sim \mathcal{N} \left(
\overline{{\bm f^*}'}, \text{cov}({\bm f^*}')
\right) \;,
\end{equation}
where
\be
\overline{{\bm f^*}'} = \text{\boldmath $\mu^*$}' + K'(\bm X^*,\bm X)
\left[K(\bm X,\bm X) + C\right]^{-1} ({\bm y}-\text{\boldmath $\mu$})
\ee
\begin{equation}
\text{cov}({\bm f^*}') = K''(\bm X^*,\bm X^*) - K'(\bm X^*,\bm X)\left[K(\bm
  X,\bm X) + C \right]^{-1}  K'(\bm X,\bm X^*) 
\end{equation}

The hyperparameters are trained in the same way as for the
reconstruction of $f(x)$ since the marginal likelihood
\eqref{log-marginal-p} that has to be maximized only depends on the
observations, but not on the function we want to reconstruct.

\subsection{Combining $f(x)$ and its derivatives}
Often we are not only interested in reconstructing $f(x)$ and its
derivatives, but also in calculating functions $g(f(x),f'(x),\dots)$,
which depend on the function derived by the Gaussian process.
Then we need to know the covariances between $f^*=f(x^*)$,
$f^*{}'=f'(x^*)$ $\dots$ at each point $x^*$ where $g$ is to be
reconstructed. These covariances are given by:
\begin{equation}
\text{cov}(f^{*(i)},f^{*(j)}) = k^{(i,j)}(x^*,x^*) - K^{(i)}(x^*,\bm X)
\left[K(\bm X,\bm X)+C\right]^{-1} K^{(j)}(\bm X,x^*)\,,\nonumber
\end{equation}
where $f^{*(i)}$ is the $i$th derivative of
$f^*$. $k^{(i,j)}(x^*,x^*)$ means that $k(x^*,x^*)$ is derived $i$
times with respect to the first argument and $j$ times with respect to
the second argument.

At each point $x^*$, $g^*=g(f^*, f^*{}' \dots)$ is then determined by
Monte Carlo sampling, where in each step $f^*$, $f^*{}'$ $\dots$ are
drawn from a multivariate normal distribution:
\begin{equation}
\begin{bmatrix} {f^*} \\ {f^*{}'} \\ \vdots \end{bmatrix} \sim
\mathcal{N} \left(
\begin{bmatrix}
\bar{f^*}\\
\bar{f^*{}'}\\
\vdots
\end{bmatrix}
,
\begin{bmatrix}
\text{var}(f^*)       & \text{cov}(f^*,f^*{}') & \cdots \\
\text{cov}(f^*,f^*{}')  & \text{var}(f^*{}') & \cdots \\
\vdots                & \vdots              & \ddots
\end{bmatrix}
\right) \,.
\end{equation}

Instead of Monte Carlo sampling, one might use propagation of errors to
determine the confidence levels of $g(f(x),f'(x),\dots)$. This is
however a first order approximation and is thus only recommended if
the errors are small, which is usually not the case when higher
derivatives are involved.

\subsection{Performance of the reconstruction}\label{performance}
Theoretically, the true function value at a point $x^*$ lies between
the 1-$\sigma$ limits of the reconstructed function (i.e. between
$f(x^*)-\sqrt{\text{Var}(x^*)}$ and $f(x^*)+\sqrt{\text{Var}(x^*)}$)
with a probability of 68\%. Thus, when reconstructing a function over
an interval in $x$ direction, one would expect the true function to
lie between the 1-$\sigma$ limits within 68\% of that interval range.
Note that this is only the expectation value. For one specific
reconstruction this percentage might be higher or lower.

\begin{figure}
\includegraphics*[width=0.4\textwidth]{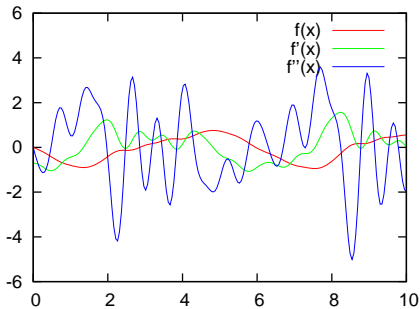}
\caption{$f^{(10)}(x)$ from equation \eqref{sin-waves} and its first
  and second derivative.}
\label{fn10}
\end{figure}

In this section, we analyse how this percentage depends on the
function that we want to reconstruct. It is a reasonable assumption
that -- given the same amount of data -- functions that change very
rapidly are more difficult to reconstruct than smooth functions. In
order to test this assumption we consider different superpositions of
sin-waves: 
\begin{equation}\label{sin-waves}
f^{(N)}(x) = \sum_i^N \frac{a_i}{b_i^2}\sin(b_ix)\;,
\end{equation}
where $N$ is the number of superpositions.
The $a_i$ are random numbers between 0.5 and 1, and with random
sign. The $b_i$ are approximately equal to $i$. (We have not used the
exact equality, because that would have made it easier for the
Gaussian Process to recognise the frequencies.) We have chosen functions in which the high frequency terms are suppressed compared to the low frequency ones as this provides the most difficult test for a data smoothing techniques ability to reconstruct derivatives, and is closest to our problem at hand. 
While in the second derivative of $f^{(N)}(x)$ the amplitudes of the
different frequencies have the same order of magnitude, higher
frequencies are suppressed in the function itself and (to a lesser
degree) in the first derivative. This is shown in Fig.~\ref{fn10}
for $N=10$. The function is much smoother than its derivatives.

\begin{figure*}
\subfloat[$f^{(N)}(x)$]{
  \includegraphics*[width=0.4\textwidth]{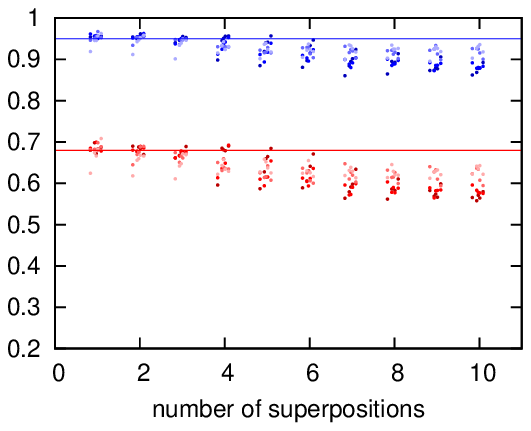}} \hspace{1cm}
\subfloat[$f^{(N)}{}'(x)$]{
  \includegraphics*[width=0.4\textwidth]{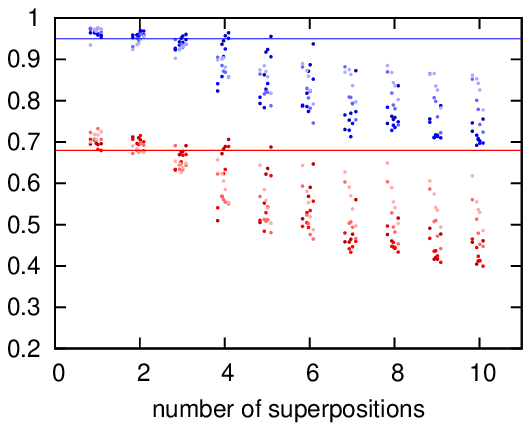}} \\ 
\subfloat[$f^{(N)}{}''(x)$]{
  \includegraphics*[width=0.4\textwidth]{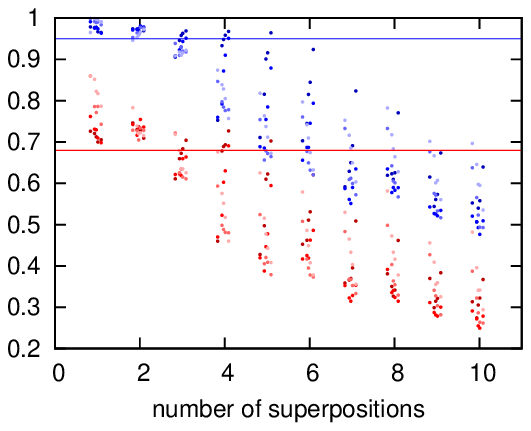}} \hspace{1cm}
\subfloat{
  \includegraphics*[width=0.4\textwidth]{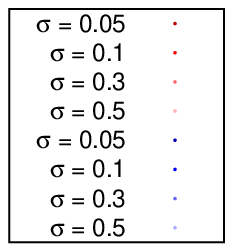}}
\caption{The function $f^{(N)}(x)$ from Eq.~\eqref{sin-waves} and its
  first and second derivatives are reconstructed using a GP, for
  different numbers of superpositions of sine waves $N$. We show the
  fractions of the range interval where the true function (top left)
  and its derivatives (first, top right; second, bottom left) a lie
  between the expected 1- and 2-$\sigma$ limits (1-$\sigma$: red
  points, 2-$\sigma$: blue points; the red and blue lines are the
  respective expectation values). Each point is the result of
  averaging over 200 mock data sets. Data sets with a smaller number
  of data are slightly shifted to the left, those with larger numbers
  to the right.}
\label{perf_test}
\end{figure*}

We have then repeatedly created mock data scattered around
$f^{(N)}(x)$, $N=1-10$, with varying noise ($\sigma=0.05, 0.1, 0.3,
0.5$) and varying number of data ($n=20, 50, 70, 100, 200$). From each
mock data set, we have reconstructed the function and its derivatives
using Gaussian Processes. We then determined the fractions of the
range interval, where the true functions $f^{(N)}(x)$, $f^{(N)}{}'(x)$ and
$f^{(N)}{}''(x)$ lie between the 1$\sigma$ and 2$\sigma$ limits of their
respective reconstructions. The result is shown in
Fig.~\ref{perf_test}. Each point corresponds to the average over 200 
realizations of the mock data set.

The reconstruction of $f^{(N)}(x)$ gives results that are quite close
to the expected values. However, when high frequencies are
involved, the errors of the reconstructed function are slightly
underestimated, leading to a smaller fraction of the true function
lying between the reconstructed 1$\sigma$ and 2$\sigma$ limits. When
the Gaussian Process fails to recognise high frequencies, this only
has a small effect on the reconstruction of $f^{(N)}(x)$, as the high
frequencies are suppressed. However, this effect becomes large for the
derivatives $f^{(N)}{}'(x)$ and $f^{(N)}{}''(x)$. 

We could be faced with this problem when reconstructing $w(z)$. If
$w(z)$ had high frequency contributions they would be suppressed in the
luminosity distance because one has to integrate twice to obtain
$d_L(z)$ from $w(z)$ using equations \eqref{lum_d} and \eqref{hubble}.

On the other hand, the errors in $f^{(N)}{}'(x)$ and $f^{(N)}{}''(x)$
are overestimated in the absence of high frequency terms.
Thus, the errors of the reconstruction reflect our lack of knowledge
about the frequencies that contribute to the differentiated
function. Oscillations that are present in the derivatives are
in general smoothed by integration and therefore hard to detect in the
function. The Gaussian process accounts for that ignorance by choosing
a balance between the cases where high frequencies are present or
absent. 

This balance is achieved by a weighted average over function space, where
more weight is on smoother functions. These represent simpler models,
which are preferred in Bayesian analysis. This can be seen when we 
perform the average over the vector of function values $\bm f^*$
(see Eq.~\eqref{f_average}). These function values are not independent
from each other, but are linked by covariances. When we consider two
points $x_1^*$ and $x_2^*$, then similar values of $f(x_1^*)$
and $f(x_2^*)$ are preferred, unless observational evidence indicates
a difference in their values. Consequently, the smoothest functions
that are consistent with the observations are preferred to functions
with higher frequencies.

\section{Reconstruction of $H(z)$, $q(z)$ and  $w(z)$}
\label{reconstruction}

There are many functions of the distance data which provide
information about dark energy dynamics. We use, in addition to the
distance and its first and second derivatives, the Hubble rate, $H(z)$,
the deceleration parameter $q(z)$, and $w(z)$. We assume $\Omega_k=0$
here so $H(z)$ provides information about the dark energy density
without the degeneracy with $\Omega_m$. Similarly, $q(z)$ provides
information about deviations from $\Lambda$CDM without this degeneracy
too.

\subsection{Mock data}
\label{subsec:DES}
We shall now use the GP method to smooth over a mock SNIa catalogue,
and reconstruct the functions $H(z)$, $q(z)$ and $w(z)$ for a given
dark energy model. As we are using the dimensionless distance $D(z)$,
the analysis of the mock data set does not depend on the present
Hubble rate $H_0$.

The Dark Energy Survey (DES) - Supernova Survey \cite{Bernstein} is
expected to obtain high quality light-curves for about 4000 SNe Ia up
to redshift $z=1.2$ in the next five years. We use the anticipated
redshift histogram given in \cite{Bernstein}, to create mock data sets
for different dark energy models: $\Lambda$CDM and a model with slowly
evolving $w(z)$. The aim of this section is to determine if the
Gaussian process can recover the correct behaviours of the respective
models.

In the following, we will assume a flat universe, $\Omega_k=0$. Then
the Hubble rate is given by
\be
H(z) = \frac{H_0}{D'(z)}
\label{hubblerec}
\ee
and the deceleration parameter by
\be
q(z) = - \frac{D''(z)}{D'(z)} (1+z) - 1 \;.
\label{qrec}
\ee
For the reconstruction of $w(z)$, we set $\Omega_m=0.3$, which is the
same matter density used to create the mock data sets.

\subsubsection{$\Lambda$CDM}

\begin{figure}
  \includegraphics[width=0.5\textwidth]{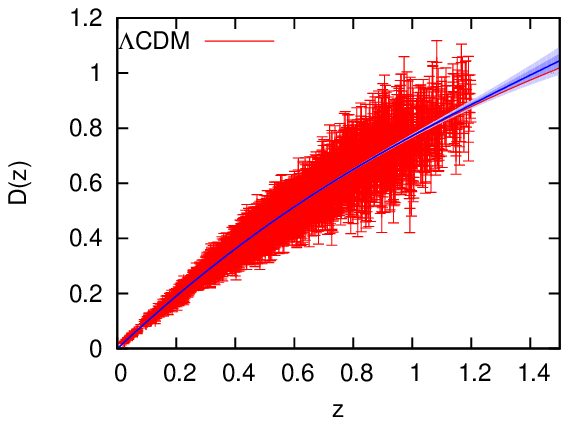}
  \includegraphics[width=0.5\textwidth]{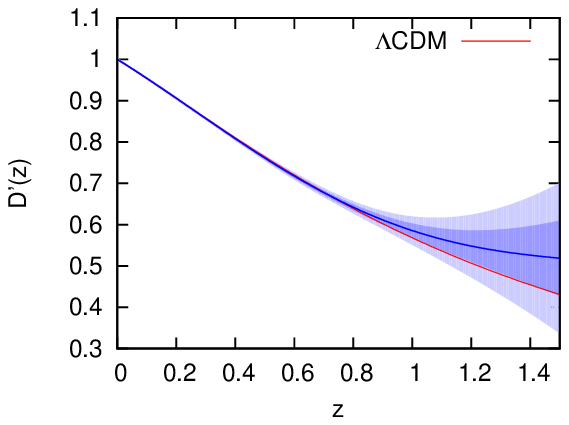}
  \includegraphics[width=0.5\textwidth]{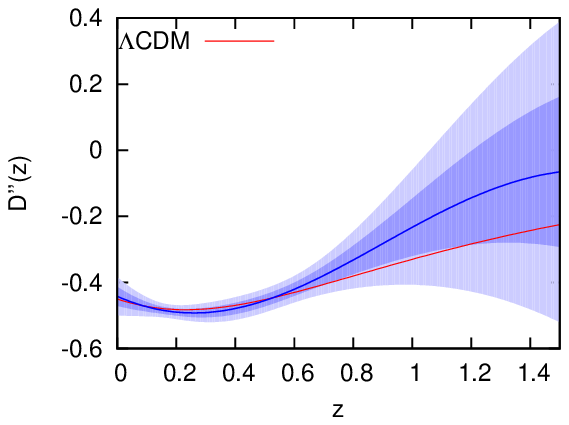}
\caption{Reconstruction of $D(z)$, $D'(z)$ and $D''(z)$ obtained from
  a mock data set following DES specifications and assuming a
  $\Lambda$CDM model with $\Omega_m=0.3$ (red line). The shaded blue
  regions are the 68\% and 95\% CL of the reconstruction.}
\label{DES_D}
\end{figure}

We start with a $\Lambda$CDM model with $\Omega_m=0.3$. We created a
mock data set according to anticipated results from DES and used Gaussian
processes (with $\mu(z)=0$) to reconstruct the distance $D(z)$ and its
derivatives. The result is shown in Fig.~\ref{DES_D}. The blue line
shows the mean of the reconstruction and the shaded areas its
68\% and 95\% CL. $D(z)$ is reconstructed with very high
precision within the redshift range of the data. At higher redshifts
the errors on the reconstruction increase, which is an expected
behaviour. For the first derivative, the point where the errors start
to increase significantly is shifted to lower redshifts.

The distance-redshift relation $D(z)$ of the $\Lambda$CDM model (red
line) and its derivatives are reconstructed nicely by the Gaussian
process.

\begin{figure}
\includegraphics*[width=0.5\textwidth]{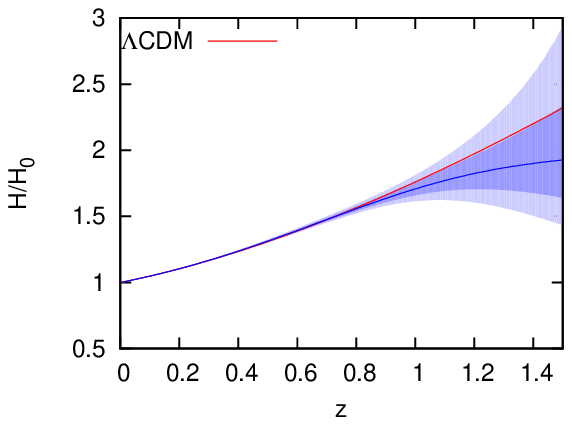}\hfill
\includegraphics*[width=0.5\textwidth]{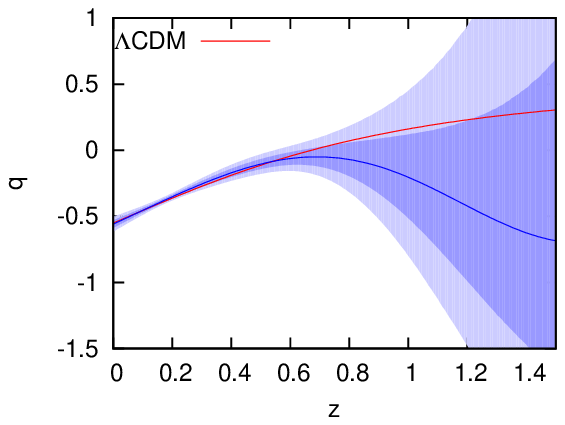}\\
\includegraphics*[width=0.5\textwidth]{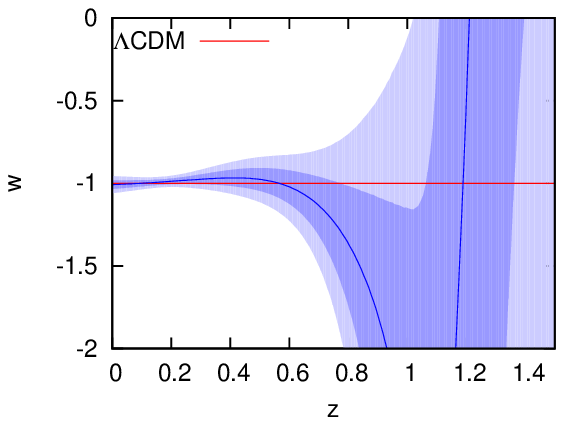}\hfill
\includegraphics*[width=0.5\textwidth]{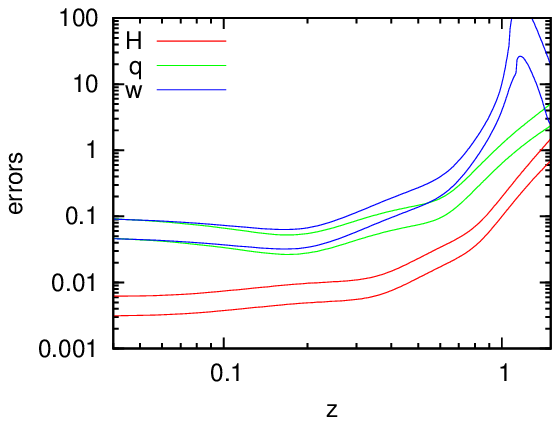}
\caption{Reconstruction of $H(z)$ (top left), $q(z)$ (top right) and $w(z)$
  (bottom left) obtained from 
  a mock data set following DES specifications and assuming a
  $\Lambda$CDM model with $\Omega_m=0.3$ (red line). The shaded blue
  regions are the 68\% and 95\% CL of the
  reconstruction. The errors on $w(z)$ correctly blow up past $z\sim1$
  where there is limited data, and the effect of $w$ on distances is
  suppressed. The bottom right plot shows the widths of the 68\% and 95\% CL
  regions.}
\label{DES_q}\label{DES_H}\label{DES_w}
\end{figure}

As the Hubble rate $H(z)/H_0$ is simply the inverse of $D'(z)$ (when
assuming $\Omega_k=0$), its reconstruction is also very precise. This
is shown in Fig.~\ref{DES_H}.

The formula for the reconstruction of $q(z)$ is slightly more complex
and contains the first and second derivatives of the distance [see
Eq.~\eqref{qrec}]. This leads to larger errors at high redshifts.
At low redshifts however, we get tight error bars~-- see Fig.~\ref{DES_q}. 

The reconstruction of the equation of state $w(z)$ requires a more
complicated formula \eqref{w}. Especially when the true value of the
denominator is close to zero, small errors in the distance can lead to
large errors in $w(z)$. In fact it can be seen in Fig.~\ref{DES_w}
that the reconstruction errors explode at redshifts $z\gtrsim 0.7$.
While the true model lies within the reconstructed 95\% CL,
a large variety of evolving dark energy models would also be
consistent with this reconstruction.

\subsubsection{Evolving $w$}

\begin{figure}
\includegraphics*[width=0.5\textwidth]{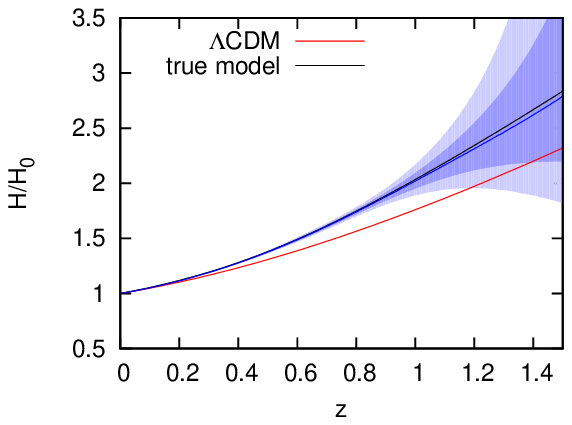}\hfill
\includegraphics*[width=0.5\textwidth]{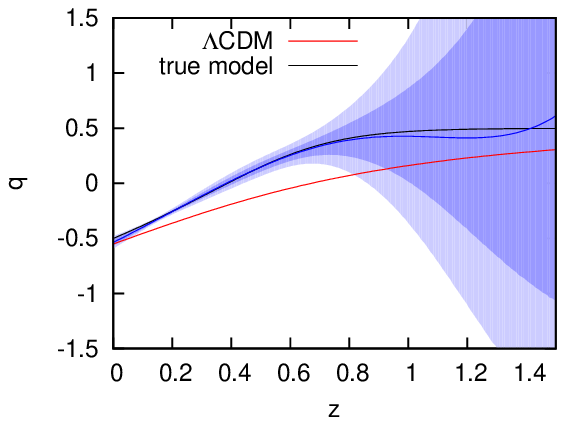}\\
\includegraphics*[width=0.5\textwidth]{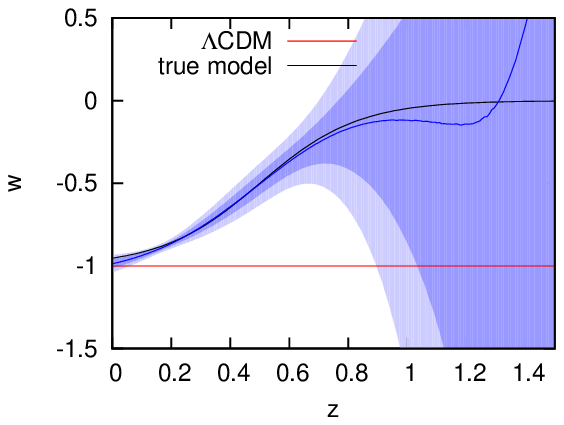}\hfill
\includegraphics*[width=0.5\textwidth]{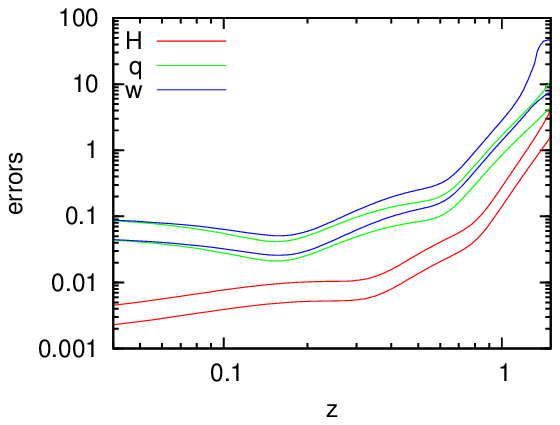}
\caption{Reconstruction of $H(z)$, $q(z)$ and $w(z)$ obtained from
  a mock data set following DES specifications and assuming a
  model with $w(z) = 0.5 (-1 + \tanh[3(z-0.5)])$ and $\Omega_m=0.3$
  (red line). The shaded blue regions are the 68\% and 95\% CL
   of the reconstruction. The bottom right plot shows the widths of
   the 68\% and 95\% CL regions.} 
\label{DES_H2}\label{DES_w2}\label{DES_q2}
\end{figure}

Next, we test a model with a slowly evolving equation of state:
\be
w(z) = \frac{1}{2} \left(-1 +
\tanh\left[3\left(z-\frac{1}{2}\right)\right]\right) \;.
\ee
The results are shown in Fig.~\ref{DES_H2}. The
Gaussian process  capture the model (black line) correctly for
$H(z)$, $q(z)$ and $w(z)$. Also shown is a $\Lambda$CDM model, which
is not consistent with the reconstruction. Thus, the Gaussian process
can correctly distinguish between these two models, assuming that we
know the matter density accurately enough.

\subsection{Union2.1 SNIa data}

In this section, we apply the Gaussian process to real SN Ia data  We
use the Union2.1 data set \cite{Suzuki}, which contains 580 SNe Ia.
We transformed the distance modulus $m-M$ given in the data set to
$D$ using
\be
m-M + 5\log\left(\frac{H_0}{c}\right) -25 = 5\log((1+z)D)
\ee
with $H_0=70$km/(s Mpc). Note that the values of $D$ do not depend on
$H_0$ itself, but on a combination of $H_0$ and the absolute magnitude
$M$, which can be written as $\mathcal{M} = M -
5\log\left({H_0}/{c}\right) + 25$. 
To account for various systematics in the data set -- such as the
calibration of the absolute magnitude -- we use the full covariance
matrix for the data. The Gaussian process analysis implicitly assumes
that the errors in $D$ follow a Gaussian distribution. However,
(relatively) small deviations from Gaussianity do not affect the
result of the Gaussian process significantly.

\begin{figure}
  \includegraphics*[width=0.5\textwidth]{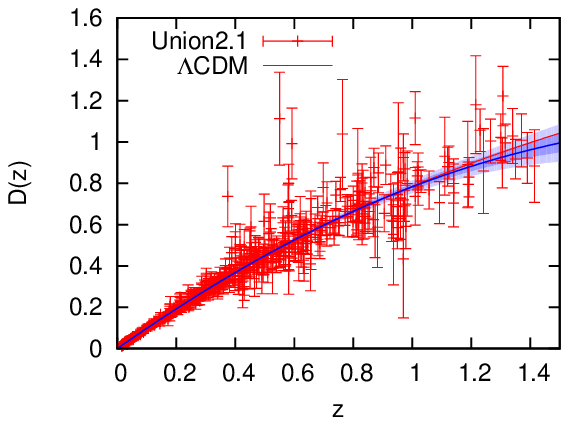}
  \includegraphics*[width=0.5\textwidth]{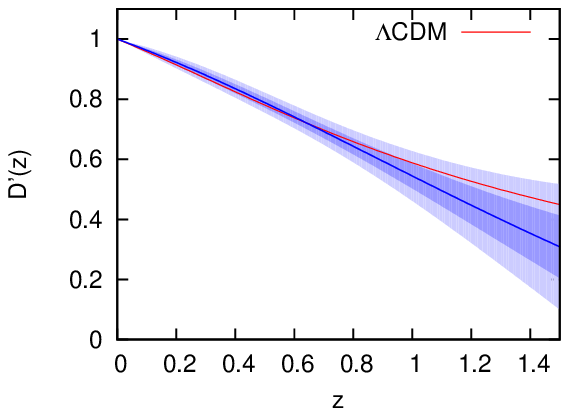}
  \includegraphics*[width=0.5\textwidth]{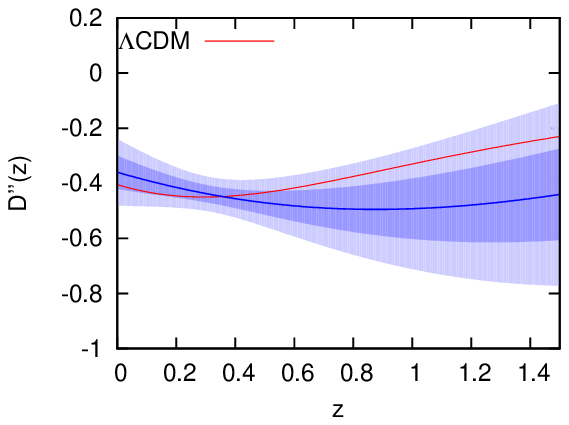}
\caption{Reconstruction of $D(z)$, $D'(z)$ and $D''(z)$ obtained from
  the Union2.1 SN data set. The shaded blue
  regions are the 68\% and 95\% CL of the
  reconstruction.
  Also shown is a $\Lambda$CDM model with $\Omega_m=0.27$.}
\label{Union_D}
\end{figure}

\begin{figure}
\includegraphics*[width=0.5\textwidth]{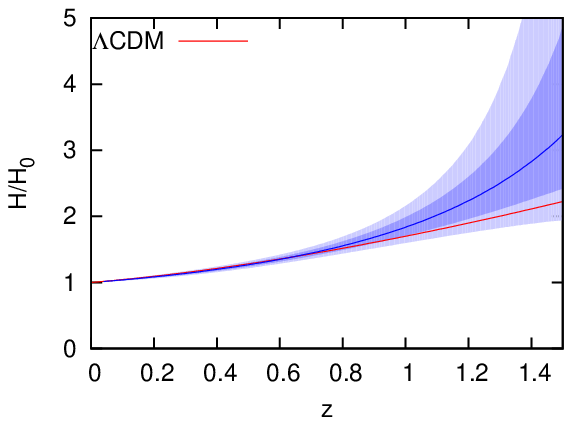}
\includegraphics*[width=0.5\textwidth]{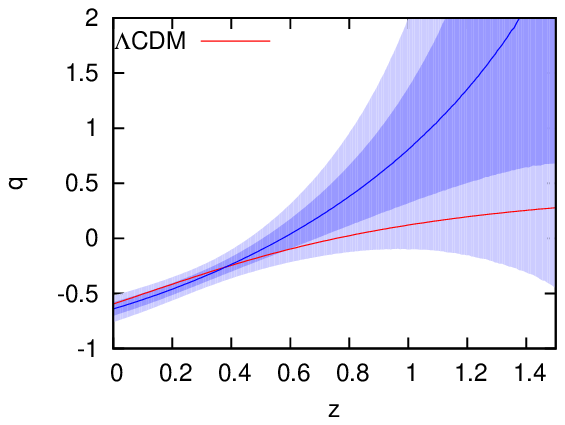}
\includegraphics*[width=0.5\textwidth]{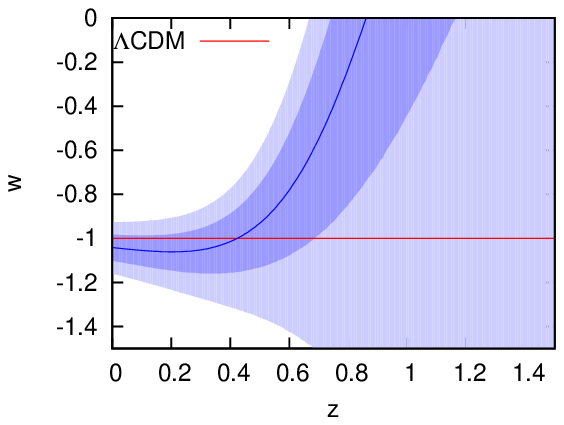}
\caption{Reconstruction of $H(z)$, $q(z)$ and $w(z)$ obtained from
  the Union2.1 SN data set. The shaded blue
  regions are the 68\% and 95\% CL of the
  reconstruction. Also shown is a $\Lambda$CDM model with
  $\Omega_m=0.27$. The reconstruction of $w(z)$ is obtained assuming
  $\Omega_m = 0.27\pm 0.015$ (constraints taken from \cite{Suzuki}).} 
\label{Union_H}\label{Union_w}\label{Union_q}
\end{figure}

The reconstruction of the distance is shown in Fig.~\ref{Union_D}. 
As expected the errors of the reconstruction are larger than
determined in~\ref{subsec:DES} for the upcoming 
DES survey due the smaller number of SNe Ia and larger measurement errors.
The reconstructions for the distance, as well as for $H(z)$, $q(z)$
and $w(z)$ (cf. Fig.~\ref{Union_q}) are consistent with $\Lambda$CDM.
For the reconstruction of $w(z)$ we have assumed $\Omega_m = 0.270\pm
0.015$, which is the constraint on the matter density for a flat
universe with time dependent equation of state given in
\cite{Suzuki}. 

\section{Discussion}\label{discussion}

We have presented a new approach to reconstructing the dark energy
equation of state using Gaussian processes. We use the GP to smooth
the data, and to produce estimates of the first and second derivatives
of the distance data. This is then combined to give $w(z)$, and any
other function of the background cosmology we might be interested
in~-- we have considered $H(z)$ and $q(z)$. This approach performs
extremely well at low and moderate redshift~-- i.e., when dark energy
affects the global cosmological dynamics (recall we do not assume a
CMB distance prior). We have shown that DES can recover $H(z)$ to
sub-percent accuracy, and $w(z)$ to a few percent. Larger errors at
high redshift simply reflect a lack of data there.

We have applied our analysis also to the Union2.1 supernova set
\cite{Suzuki}. The results are consistent with $\Lambda$CDM. But note
that the distance moduli in this data set were derived in a
model dependent way; the cosmological model is fitted at the same time
as some of the nuisance parameters. For a fully self-consistent
analysis, we would need to include the derivation of the distance
moduli directly into our Gaussian process analysis. This is beyond the
scope of the present work. While it would certainly be interesting to
perform a more realistic analysis in future work, the present paper
already shows that Gaussian processes are a powerful analysis tool.

Yet, one has to be careful when interpreting the results. As we have
shown in section~\ref{performance}, the goodness of the reconstruction
depends on the smoothness of the true function and the quality of the
data. High frequency terms in $w(z)$ are very hard to detect by
distance measurements. If such terms were present, we would not be
able to see them with the present Union2.1 SNe, nor with the future
DES SNe. A slowly evolving equation of state is however often captured
within the 68\% CL, i.e the errors are overestimated. The
reconstructed errors represent our lack of knowledge about the
smoothness of the function. The Gaussian process automatically
determines the errors such that a balance between very smooth
functions and rapidly oscillating functions is obtained.

While we started with the reconstruction of $D(z)$ using Gaussian
Processes and subsequently determined $w(z)$, Holsclaw {\em et al.}
\cite{Holsclaw10} followed a different approach. They use a GP-based
MCMC algorithm. Starting with initial values for the hyperparameters
and for the vector $\bm w$ (containing values of $w(z)$ at different
redshifts), they perform the first integral over the Gaussian Process
of $\bm w$ analytically and the second one numerically, and finally
calculate the distance modulus. The hyperparameters and $\bm w$ are then
varied in each step of the MCMC and the resulting distance
modulus is compared to observational data from the Constitution set
\cite{Hicken}.

Using this method, the errors on $w(z)$ are much more uniform than in
our approach. Our errors are smaller at low redshifts and larger at
high redshifts. In~\cite{Holsclaw10}, the errors do not depend
strongly on redshift -- even at high redshifts, where the data density
is significantly smaller, and the effect of dark energy on the
cosmological dynamics very weak. In fact, their errors on $w$ are
smaller than those induced from uncertainties in $\Omega_m$ (note that
they use broader priors than we have assumed in the left panel of
Fig.~\ref{exactD}). Their method is optimised towards smooth $w(z)$;
this preference is much stronger than in our approach, which prefers
smooth distances instead. This is similar for other approaches which
focus on $w(z)$ (see e.g.,~\cite{Crittenden:2011aa}). Note, however,
that we use different data assumptions, and do not assume a CMB
distance prior.

Of course, methods which start form $w$ provide valuable constraints
on dark energy, and we do not advocate reconstruction as a replacement
of such approaches. Instead they should be considered as complementary
as it helps us understand how differing priors used in the
construction of $w(z)$ affects our final result. In addition, by not
assuming that the dark energy equation of state has physical
significance, a reconstruction approach allows us to consider more
general models where the effective $w(z)$ is ill defined. Constraints
on the expansion dynamics are readily obtained without invoking dark
energy models at all. Furthermore it is readily used for non-standard
models, and consistency tests of the FLRW models themselves, as we
shall consider in future work.

\acknowledgments

We thank Phil Bull for suggesting an improved way to calculate the
errors of $q$, $w$. We also thank Rob Crittenden, J\"o Fahlke, Alan
Heavens, Roy Maartens, Roberto Trotta, Melvin Varughese and Sahba Yahya
for discussions.\\
This work is funded by the NRF (South Africa).\\
\\
An independent study \cite{Shafieloo12} of some aspects of the work
presented here was placed on the arXiv within a few days of this work.

\appendix
\section{Conditional distribution of the multivariate normal distribution}

Starting from the joint distribution for $\bm y$ and $\bm f^*$, we
want to calculate the conditional distribution $\bm f^*|\bm y$. The
joint distribution is given by
\begin{equation}
\begin{bmatrix} {\bm y} \\ {\bm f^*} \end{bmatrix} \sim
\mathcal{N} \left( 
\begin{bmatrix}
\text{\boldmath $\mu$}\\
\text{\boldmath $\mu^*$}
\end{bmatrix}
,
\begin{bmatrix}
\tilde{K} & K^*\\
K^{*T}     & K^{**}
\end{bmatrix}
\right)
\end{equation}
where we have used the abbreviations $\tilde{K} = K(\bm X,\bm X) +C$,
$K^* = K(\bm X,\bm X^*)$ and $K^{**} = K(\bm X^*,\bm X^*)$. Denoting
the combined covariance matrix as $K$, the joint probability
distribution is: 
\begin{equation} \label{joint}
p(\bm y, \bm f^*) = \frac{1}{(2\pi)^{N/2} \sqrt{\det(K)}}
\exp\left\{ -\frac{1}{2}
\left[ (\bm y - \bm \mu)^T, (\bm f^* - \bm \mu^*)^T\right]
K^{-1}
\begin{bmatrix}
\bm y - \bm \mu\\
\bm f^* - \bm \mu^*
\end{bmatrix}
\right\} 
\end{equation}
with $N=n+n^*$, where $n$ and $n^*$ are the dimensions of $\bm y$ and
$\bm f^*$, respectively.

The determinant for block matrices is given by
\begin{equation}
\det(K) = \det(\tilde{K}) \det(K^{**} - K^{*T} K^{**-1}K^*)
\end{equation}
and the inverse by
\begin{equation}
K^{-1} =
\begin{bmatrix}
 M_{11} & M_{12}\\
 M_{21} & M_{22}
\end{bmatrix}
\end{equation}
where
\begin{eqnarray}
M_{11} &=& (\tilde{K} - K^* K^{**-1} K^{*T})^{-1} \\
M_{22} &=& (K^{**} - K^{*T}\tilde{K}^{-1}K^*)^{-1} \\
M_{12} &=& -\tilde{K}^{-1} K^* (K^{**} - K^{*T}\tilde{K}^{-1}K^*)^{-1}
= M_{21}^T
\end{eqnarray}
Using the matrix inversion lemma, we can write $M_{11}$ as
\be
M_{11} = \tilde{K}^{-1} + \tilde{K}^{-1} K^* (K^{**} - K^{*T}
\tilde{K}^{-1} K^*)^{-1} K^{*T} \tilde{K}^{-1}
\ee

Inserting these results into equation \eqref{joint}, we get for the
joint probability distribution
\begin{eqnarray}
p(\bm y, \bm f^*) &=&
\frac{1}{(2\pi)^{n/2} \sqrt{\det(\tilde{K})}}
\exp\left[ -\frac{1}{2} (\bm y - \bm \mu)^T \tilde{K}^{-1}
(\bm y - \bm \mu) \right] \nonumber\\
&&{}+ \frac{1}{(2\pi)^{n^*/2} \sqrt{\det(A)}}
\exp\left[ -\frac{1}{2} (\bm f^* - \bm b)^T A^{-1}
(\bm f^* - \bm b) \right] 
\end{eqnarray}
with
\begin{eqnarray}
\bm b &=& \bm \mu^* + K^{*T} \tilde{K}^{-1} (\bm y - \bm \mu) \\
A &=& K^{**} - K^{*T} \tilde{K}^{-1} K^*
\end{eqnarray}

The marginal probability distribution of $\bm y$ is
\begin{eqnarray}\label{f_average}
p(\bm y) &=& \int p(\bm y, \bm f^*) d\bm f^* \\
&=& \frac{1}{(2\pi)^{n/2} \sqrt{\det(\tilde{K})}}
\exp\left[ -\frac{1}{2} (\bm y - \bm \mu)^T \tilde{K}^{-1}
(\bm y - \bm \mu) \right]\nonumber
\end{eqnarray}
and the conditional probability distribution
\begin{eqnarray}
p(\bm f^*|\bm y) &=& \frac{p(\bm y, \bm f^*)}{p(\bm y)}\nonumber\\
&=&\frac{1}{(2\pi)^{n^*/2} \sqrt{\det(A)}}
\exp\left[ -\frac{1}{2} (\bm f^* - \bm b)^T A^{-1}
(\bm f^* - \bm b) \right] \nonumber\\
&=& \mathcal{N}(\bm b,A)
\end{eqnarray}
Note that $\bm b$ and $A$ are equal to $\overline{\bm f^*}$ and
$\text{cov}(\bm f^*)$ of equation \eqref{post}.

\end{document}